\documentclass[11pt,preprint]{aastex}

\begin{document}

\shorttitle{Dust and CO in the TW Hya Disk}

\shortauthors{Andrews et al.}

\title{The TW Hya Disk at 870\,$\mu$m: Comparison of CO and Dust Radial Structures}

\author{Sean M. Andrews\altaffilmark{1}, David J. Wilner\altaffilmark{1}, A. M. Hughes\altaffilmark{2}, Chunhua Qi\altaffilmark{1}, Katherine A. Rosenfeld\altaffilmark{1}, Karin I. {\"{O}}berg\altaffilmark{1,3}, T. Birnstiel\altaffilmark{4}, Catherine Espaillat\altaffilmark{1,5}, Lucas A. Cieza\altaffilmark{6}, Jonathan P. Williams\altaffilmark{6}, Shin-Yi Lin\altaffilmark{7}, and Paul T. P. Ho\altaffilmark{1,8}}

\altaffiltext{1}{Harvard-Smithsonian Center for Astrophysics, 60 Garden Street, Cambridge, MA 02138}
\altaffiltext{2}{University of California at Berkeley, Department of Astronomy, 601 Campbell Hall, Berkeley, CA 94720}
\altaffiltext{3}{Hubble Fellow}
\altaffiltext{4}{Ludwig-Maximilians-Universit{\"{a}}t, University Observatory Munich, Scheinerstrasse 1, D-81679 Munich, Germany}
\altaffiltext{5}{NSF Astronomy \& Astrophysics Postdoctoral Fellow}
\altaffiltext{6}{University of Hawaii Institute for Astronomy, 2680 Woodlawn Drive, Honolulu, HI 96822}
\altaffiltext{7}{Center for Astrophysics \& Space Science, University of California San Diego, La Jolla, CA 92093}
\altaffiltext{8}{Academia Sinica Institute of Astronomy and Astrophysics, P.O. Box 23-141, Taipei 106, Taiwan}

\begin{abstract}
We present high resolution ($0\farcs3 = 16$\,AU), high signal-to-noise ratio 
Submillimeter Array observations of the 870\,$\mu$m (345\,GHz) continuum and CO 
$J$=3$-$2 line emission from the protoplanetary disk around TW Hya.  Using 
continuum and line radiative transfer calculations, those data and the 
multiwavelength spectral energy distribution are analyzed together in the 
context of simple two-dimensional parametric disk structure models.  Under the 
assumptions of a radially invariant dust population and (vertically integrated) 
gas-to-dust mass ratio, we are unable to simultaneously reproduce the CO and 
dust observations with model structures that employ either a single, distinct 
outer boundary or a smooth (exponential) taper at large radii.  Instead, we 
find that the distribution of millimeter-sized dust grains in the TW Hya disk 
has a relatively sharp edge near 60\,AU, contrary to the CO emission (and 
optical/infrared scattered light) that extends to a much larger radius of at 
least 215\,AU.  We discuss some possible explanations for the observed radial 
distribution of millimeter-sized dust grains and the apparent CO-dust size 
discrepancy, and suggest that they may be hallmarks of substructure in the dust 
disk or natural signatures of the growth and radial drift of solids that might 
be expected for disks around older pre-main sequence stars like TW Hya.  \\
\end{abstract}

\section{Introduction}

The physical conditions of the gas and dust in young circumstellar disks shape 
the formation and early evolution of planetary systems.  The spatial 
distribution of disk material -- the density structure -- plays many 
fundamental roles, and is especially relevant for dictating when and where 
planets can form and migrate in the disk.  Naturally, empirical constraints on 
protoplanetary disk structures are of significant value in efforts to develop 
realistic models of the complex processes involved in planet formation.  Some 
recent studies have made progress in extracting the radial (surface) density 
profiles in these disks from high angular resolution measurements of their 
millimeter-wave continuum emission 
\citep{andrews09,andrews10b,isella09,guilloteau11}.  However, such observations 
are only sensitive to the density structure of the (presumably) trace 
population of disk solids, and not the gas that wields far more influence over 
the evolution of disk structure and the formation of planets.  The major 
challenge is that the gas in these disks is primarily cold H$_2$, which is not 
directly observable.  And while there have been many investigations of less 
abundant molecular species \citep[see][]{dutrey07}, limited sensitivity to low 
optical depth lines and an incomplete understanding of the complicated 
chemistry in these disks have so far made it difficult to convert observations 
of the gas into density constraints \citep[see][]{williams11}.  

Given those obstacles, there have not been many attempts to directly compare 
the gas and dust structures of protoplanetary disks with spatially resolved 
data.  A few studies used simple models to fit the millimeter continuum (dust) 
and CO line (gas) emission {\it independently}, finding inconsistent structures 
where the dust is much more compact than the gas \citep{pietu05,isella07}.  
\citet{hughes08} suggested that this discrepancy was likely an optical depth 
illusion, an artifact of the assumed sharp outer edge in the density model.  
They showed that a surface density profile with a smooth taper at large radii 
could better reproduce the dust and gas observations.  However, \citet{panic09} 
argued that similar modifications to the model structure of the IM Lup disk 
\citep[see][]{pinte08} were insufficient to account for the much different 
sizes they measure for its gas and dust emission.  Recently, \citet{qi11} 
successfully matched their observations of multiple CO transitions and 
continuum emission from the HD 163296 disk with a single density model, 
although refinements to that model based on the detailed morphology of the 
continuum emission were not a priority in that study.  In general, there is 
still substantial uncertainty that the gas and dust trace the same structures 
in protoplanetary disks.  From a practical standpoint, that uncertainty is 
disconcerting.  Our knowledge of the mass contents of these disks is entirely 
based on their (easy to measure) dust emission, but we are not sure if those 
dust structures also describe the gas reservoirs that effectively control the 
key disk evolution and planet formation processes.  

Although direct measurements of H$_2$ densities are not possible, in principle 
an {\it indirect} comparison of the dust and gas structures in a protoplanetary 
disk can be made with sufficiently sensitive and high resolution observations 
of the thermal continuum and emission lines from trace gas species.  To meet 
those requirements, a relatively massive and nearby disk would be an ideal 
candidate target.  TW Hya is an isolated, $\sim$0.8\,M$_{\odot}$ T Tauri star 
located only $54\pm6$\,pc from the Sun 
\citep{rucinski83,wichmann98,vanleeuwen07}.  Despite its advanced age 
\citep[$\sim$10\,Myr;][]{kastner97,webb99}, it hosts a massive, gaseous 
accretion disk with a rich molecular spectrum and strong continuum emission out 
to centimeter wavelengths \citep{wilner00,wilner03,wilner05,qi04,qi06,qi08}.  
Those observations and a suite of resolved optical/infrared scattered light 
images confirm that the disk is well-resolved and viewed nearly face-on 
\citep{krist00,trilling01,weinberger02,apai04,roberge05}.  The inner edge of 
this ``transition" disk is truncated at a radius of 4\,AU, likely by an unseen 
(and maybe planetary) companion \citep{calvet02,hughes07}.  Given its 
proximity, favorable orientation, and rich gas and dust content, the TW Hya 
disk is uniquely well-suited for a comparative investigation of the radial 
behavior of its gas and dust structures.

In this article, we present new sub-arcsecond resolution observations of the 
870\,$\mu$m continuum and CO $J$=3$-$2 emission from the TW Hya disk.  Using 
state-of-the-art radiative transfer modeling tools, we use those measurements 
to compare the radial distributions of its CO gas and dust.  In \S 2 we 
describe the observations and data calibration.  The basic characteristics of 
the data are reviewed in \S 3.  A detailed description of the modeling and 
results are provided in \S 4.  The implications of that analysis are discussed 
in \S 5, and our principle conclusions are summarized in \S 6.

\section{Observations and Data Reduction}

TW Hya was observed at 345\,GHz (870\,$\mu$m) with the Submillimeter Array 
\citep[SMA;][]{ho04} on 8 occasions in 2006-2010 in the compact (C), extended 
(E), and very extended (V) array configurations, providing baseline lengths of 
16-70\,m, 28-226\,m, and 68-509\,m, respectively.  To accomodate various 
studies of the CO $J$=3$-$2 emission line (at 345.796\,GHz), the SMA double 
sideband receivers and correlator were configured with several different local 
oscillator (LO) settings and spectral resolutions throughout this campaign.  
The ``default" setup used an LO frequency of 340.755\,GHz (880\,$\mu$m) with 
the CO line in the center of the upper sideband, with a resolution of 
0.70\,km s$^{-1}$ in one spectral chunk.  The other 23 partially-overlapping 
104\,MHz spectral chunks in each sideband were sampled coarsely with 32 
channels each.  Some of the 2008 observations utilized a ``high" resolution 
correlator mode, with a large portion of the bandwidth devoted to probing the 
CO line with 0.044\,km s$^{-1}$ channels \citep[see][]{hughes11}.  While those 
data sample the continuum with the default spectral resolution, they have a 
reduced continuum bandwidth (2.6\,GHz; $\sim$70\%\ of the available bandwidth 
at the time) and a higher LO frequency at 349.935\,GHz (857\,$\mu$m).  The 2010 
observations were conducted in a ``medium" resolution mode that employed the 
new expanded bandwidth capabilities at the SMA.  In that case, two different 
2\,GHz IF bands were centered $\pm$5\,GHz (the normal setup) and $\pm$7\,GHz 
from the LO frequency, 339.853\,GHz (883\,$\mu$m).  The CO line was observed at 
a moderate resolution of 0.18\,km s$^{-1}$ in the upper sideband of the first 
IF band, and the continuum was coarsely sampled across the remaining expanded 
bandwidth.  Finally, a ``hybrid" mode was utilized at the end of 2006 to cover 
the $J$=4$-$3 transition of H$^{13}$CO$^+$ \citep[see][]{qi08}.  A summary of 
the observational setups is provided in Table \ref{obs_journal}.

\begin{deluxetable}{lccccc}
\tablecolumns{6}
\tablewidth{0pc}
\tablecaption{Summary of SMA Observations\label{obs_journal}}
\tablehead{
\colhead{UT Date} & \colhead{array} & \colhead{spectral} & \colhead{RMS noise} & \colhead{beam size} & \colhead{beam PA} \\
\colhead{}        & \colhead{config.}      & \colhead{setup}              & \colhead{[mJy beam$^{-1}$]}    & \colhead{[\arcsec]}  & \colhead{[\degr]} \\
\colhead{(1)}     & \colhead{(2)}   & \colhead{(3)}           & \colhead{(4)}      & \colhead{(5)}        & \colhead{(6)}}
\startdata
2006 Dec 28 & C   & hybrid  & 3.7     & $4.2\times1.7$ & 4     \\
2008 Jan 23 & E   & high    & 3.2     & $0.9\times0.7$ & 21    \\
2008 Feb 20 & E   & high    & 3.1     & $1.0\times0.7$ & $-$20 \\
2008 Feb 21 & E   & default & 4.1     & $1.2\times0.5$ & 14    \\
2008 Mar 2  & C   & high    & 3.6     & $3.9\times1.9$ & $-$17 \\
2008 Mar 9  & C   & default & 3.2     & $3.7\times1.8$ & $-$13 \\
2008 Apr 3  & V   & default & 2.2     & $0.6\times0.3$ & 19    \\
2010 Feb 9  & V   & medium  & 1.8     & $0.5\times0.3$ & $-$3  \\
\hline
combined    & all & \nodata & 2.0     & $0.8\times0.6$ & $-$3  \\
\enddata
\tablecomments{Col.~(1): UT date of observations.  Col.~(2): SMA array
configuration, C = compact, E = extended, and V = very extended.  Col.~(3):
Adopted correlator mode (see \S 2).  Col.~(4): Continuum RMS noise in a
naturally-weighted map.  Col.~(5): Naturally-weighted synthesized beam 
dimensions.  Col.~(6): Naturally-weighted synthesized beam orientation (major 
axis position angle, measured east of north).}
\end{deluxetable}

TW Hya was observed in an alternating sequence with the nearby quasar 
J1037-295 (a projected separation of 7.3\degr), with a total cycle time of 15 
minutes in the C and E configurations and 8-10 minutes in the V configuration.  
The quasars 3C 279 or J1146-289 were observed every other cycle.  For the 
tracks on 2008 February 21, March 9, and April 3, the target portion of the 
cycle was shared with the nearby sources HD\,98800 and Hen\,3-600 
\citep[see][]{andrews10a}.  Planets, satellites, and bright quasars were 
observed as bandpass and absolute flux calibrators when TW Hya was at low 
elevations ($<$18\degr), depending on their availability and the array 
configuration.  Most of these data were obtained in the best atmospheric 
conditions available at Mauna Kea, with zenith opacities of 0.03-0.05 at 
225\,GHz (0.6-1.0\,mm of precipitable water vapor) and well-behaved phase 
variations on timescales longer than the calibration cycle.  The conditions on 
2008 March 2 and 9 were worse, but typical for Mauna Kea (with 1.4-1.8\,mm of 
precipitable water vapor).  

The data from each individual observation were reduced independently with the 
IDL-based {\tt MIR} software package.  The bandpass response was calibrated 
with observations of bright planets and quasars, and broadband continuum 
channels in each sideband (and IF band, where applicable) were generated by 
averaging and then combining the central 82\,MHz in each line-free spectral 
chunk.  The visibility amplitude scale was derived from observations of Uranus, 
Titan, Callisto, or Vesta, with a typical systematic uncertainty of 10-15\%.  
The antenna-based complex gain response of the system was determined with 
reference to J1037-295.  The observations of 3C 279 or J1146-289 were used to 
assess the quality of phase transfer in the gain calibration process.  Based on 
those data, we estimate that the ``seeing" generated by atmospheric phase noise 
and any small baseline errors is small, 0.10-0\farcs15.  This result is a 
testament to the high quality of the observing conditions, especially given the 
low target elevation and wide projected separations between the calibrators 
(for reference, 3C 279 is located 39.7\degr\ from TW Hya and 38.5\degr\ from 
J1037-295, while the corresponding separations for J1146-289 are 11.0\degr\ and 
15.1\degr, respectively).  

Before the observations could be combined, we had to account for the proper 
motion of TW Hya over the $\sim$3 year observing baseline 
\citep[$\mu_{\alpha}\cos{\delta} = -0\farcs066$\,yr$^{-1}$, $\mu_{\delta} = 
-0\farcs014$\,yr$^{-1}$;][]{vanleeuwen07}.  Fortunately, each individual 
dataset exhibited bright, symmetric, centrally-peaked continuum emission that 
we associate with dust in the TW Hya disk.  Using elliptical Gaussian fits to 
those continuum visibilities, we determined centroid positions for each 
individual set of observations and associated them with the stellar position at 
that epoch.  The measured emission centroids are consistent with the expected 
stellar positions, well within the $\sim$0\farcs1 absolute astrometric 
accuracy of the SMA (set primarily by small baseline uncertainties).  Based on 
those centroid measurements, each individual dataset was aligned to a common 
coordinate system.  After confirming that the aligned visibility sets showed 
excellent agreement on all overlapping baselines, all datasets were combined to 
produce composite 345\,GHz continuum and CO $J$=3$-$2 visibilities.  

The composite visibilities were Fourier inverted, deconvolved with the {\tt 
CLEAN} algorithm, and restored with a synthesized beam using the {\tt MIRIAD} 
software package.  The natural weighting of the continuum data produces an 
image with a $0\farcs80\times0\farcs58$ beam and an RMS noise level of 2.0\,mJy 
beam$^{-1}$ (which is dynamic-range limited).  Composite CO $J$=3$-$2 channel 
maps were synthesized with a velocity resolution of 0.2\,km s$^{-1}$ and a 
circular beam with a FWHM of 1\farcs0 (the naturally-weighted resolution is 
$0\farcs96\times0\farcs73$).  The RMS noise level is 0.13\,Jy beam$^{-1}$ in 
each channel.  

Ancillary information in the literature was used to construct the TW Hya 
spectral energy distribution (SED) that will be used with the SMA data.  
We adopted the optical monitoring results of \citet{mekkaden98} and 
near-infrared photometry from \citet{weintraub00} and the 2MASS point source 
catalog \citep{cutri03}.  In the thermal infrared, {\it Spitzer} 
flux densities measured by \citet{hartmann05} and \citet{low05} were 
supplemented with {\it IRAS} and {\it Herschel} data \citep{weaver92,thi10}.  
We also include a {\it Spitzer} IRS spectrum, kindly provided by E.~Furlan 
\citep[see][]{uchida04}.  At submillimeter wavelengths, we relied on the 
calibrator flux densities from the SHARC-II\footnote{\url{http://www.submm.caltech.edu/$\sim$sharc/analysis/calibration.htm}} and SCUBA\footnote{\url{http://www.jach.hawaii.edu/JCMT/continuum/calibration/sens/potentialcalibrators.html}} instruments at the Caltech Submillimeter Observatory and 
James Clerk Maxwell Telescope, respectively ($F_{\nu} = 6.13\pm0.68$, 
$3.9\pm0.7$, and $1.37\pm0.01$\,Jy at 350, 443, and 869\,$\mu$m, 
respectively).  Additional millimeter-wave measurements were collected from 
\citet{weintraub89}, \citet{qi04}, and \citet{wilner03}.

\section{Results}

\begin{figure}
\epsscale{1.02}
\plotone{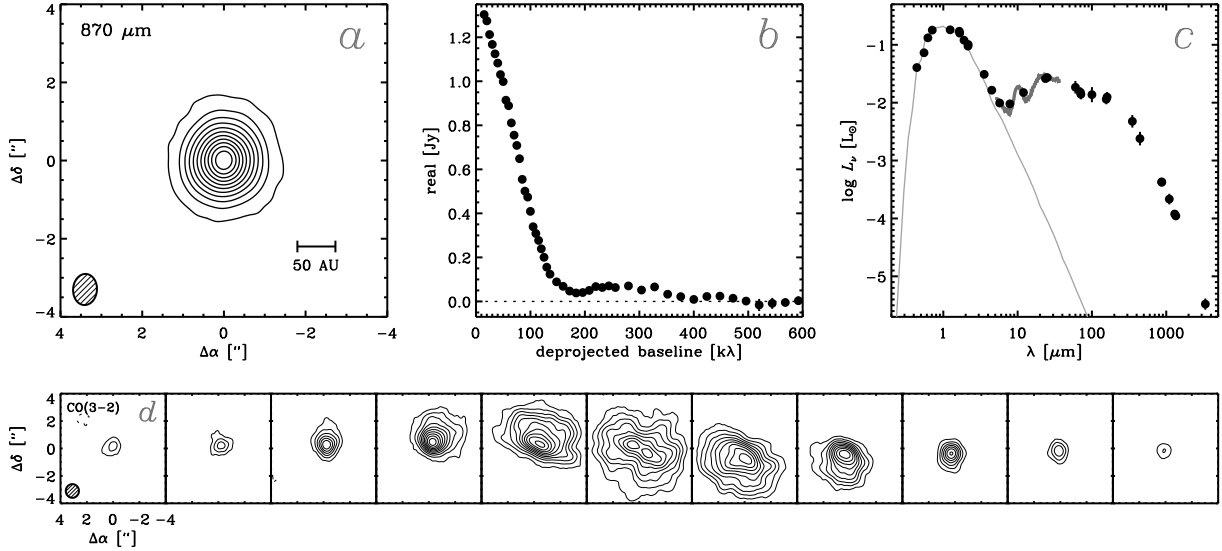}
\figcaption{($a$) Naturally weighted composite image of the 870\,$\mu$m 
continuum emission from the TW Hya disk.  Contours start at 10\,mJy beam$^{-1}$
(5\,$\sigma$) and increase in 30\,mJy beam$^{-1}$ (15\,$\sigma$) increments.  
The synthesized beam dimensions are shown in the lower left.  ($b$) Azimuthally 
averaged 870\,$\mu$m continuum visibility profile as a function of the 
deprojected baseline length (real part only; the imaginary terms are 
effectively zero on all baselines).  The uncertainties are typically smaller 
than the symbol sizes.  Note the low-amplitude oscillations beyond 
$\sim$150\,k$\lambda$.  ($c$) Complete SED for the TW Hya star+disk system 
(references are in the text; see \S 2).  The {\it Spitzer} IRS spectrum is 
shown as a thick gray curve.  Our adopted stellar photosphere model is 
overlaid as a thin gray curve (see \S 4.1).  ($d$) CO $J$=3$-$2 channel maps 
from the TW Hya disk, on the same angular scale as the continuum map in panel 
$a$.  Contours are drawn at 0.4\,Jy beam$^{-1}$ intervals ($\sim$3\,$\sigma$) 
in each 0.2\,km s$^{-1}$-wide channel.  The 1\arcsec\ synthesized beam is shown 
in the lower left.  The central channel represents the TW Hya systemic 
velocity, at $V_{\rm LSR} = +2.86$\,km s$^{-1}$.  \label{data}}
\end{figure}

The astrometrically aligned and combined SMA data are shown together in Figure 
\ref{data}, along with the broadband SED.  The synthesized continuum image in 
Figure \ref{data}$a$ has an effective frequency of 344.4\,GHz (870\,$\mu$m), 
with an integrated flux density of $1.34\pm0.13$\,Jy and a peak flux density of 
$0.337\pm0.034$\,Jy beam$^{-1}$ (a peak signal-to-noise ratio of $\sim$170), 
including the systematic calibration uncertainties.  The continuum emission is 
regular and symmetric on the angular scales probed here, with a synthesized 
beam size of $43\times31$\,AU projected on the sky.  Essentially no emission is 
detected outside a $\sim$1\arcsec\ radius.  Given that the integrated flux 
density determined from the SMA data is in good agreement with single-dish 
measurements \citep[see \S 2;][]{weintraub89,difrancesco08}, it is clear that 
this is no optical depth effect: {\it all} of the millimeter-wave dust emission 
from the TW Hya disk is concentrated inside a projected radius of $\sim$60\,AU. 

While the continuum image appears rather plain, some interesting emission 
features are apparent in a direct examination of the visibilities.  Figure 
\ref{data}$b$ displays the azimuthally averaged 870\,$\mu$m continuum real 
visibilities as a function of the deprojected baseline length 
\citep[see][]{andrews09}, assuming the inclination ($i = 6\degr$) and major 
axis position angle (PA = 335\degr) derived from high spectral resolution CO 
emission line data by \citet{hughes11}.  The imaginary visibilities are 
effectively zero within the noise on all sampled baselines, confirming the 
accuracy of the astrometric alignment and reinforcing that there are no obvious 
departures from axisymmetry in the TW Hya disk.  The visibility profile in 
Figure \ref{data}$b$ shows a smooth decrease out to $\sim$150\,k$\lambda$, 
followed by low-amplitude oscillations on longer baselines and an apparent null 
near 500\,k$\lambda$.  These features are distinct in independent datasets 
(particularly the ``dip" near 180\,k$\lambda$, where 4 different observations 
span that range of baselines), are present regardless of the bin sizes used for 
profile averaging, and are not noted in the visibility profiles for the test 
calibrators (J1146-289 or 3C 279).  Despite the challenges of calibrating SMA 
data for low-elevation targets, the persistence of these visibility modulations 
make us confident that they are real features.  Moreover, we will demonstrate 
in \S 4 that they can be reproduced with models that incorporate a sharp outer 
edge in their emission profiles.  The null at 500\,k$\lambda$ is consistent 
with the 4\,AU-radius inner disk cavity inferred from the SED \citep[see Figure 
\ref{data}$c$;][]{calvet02} and a VLA 7\,mm continuum image \citep{hughes07}.  

The panels in Figure \ref{data}$d$ show the composite CO $J$=3$-$2 emission 
line channel maps for the TW Hya disk, resampled to a velocity resolution of 
0.2\,km s$^{-1}$ with a circular 1\farcs0 (54\,AU) synthesized beam.  The 
emission is firmly detected ($>$3\,$\sigma$) out to $\pm$1.2\,km s$^{-1}$ from 
the systemic velocity ($V_{\rm LSR} = 2.86$\,km s$^{-1}$), with an integrated 
intensity of $34.8\pm3.5$\,Jy km s$^{-1}$ and a peak flux of $4.2\pm0.4$\,Jy 
beam$^{-1}$ ($43\pm4$\,K), including the calibration uncertainties.  Those 
values are in good agreement with previous single-dish and SMA measurements 
\citep{vanzadelhoff01,qi04,hughes11}.  The channel maps in Figure \ref{data}$d$ 
show a clear rotation pattern, from northwest (blueshifted) to southeast 
(redshifted), with a narrow line-width due to a face-on viewing geometry.  Near 
the systemic velocity, the CO emission subtends $\sim$4\arcsec\ (215\,AU) in 
radius.

\section{Modeling Analysis}

These SMA observations offer some new insights into the TW Hya disk structure.  
Naturally, as one of the nearest pre-main sequence stars, TW Hya and its 
associated disk have been the subject of intense observational scrutiny.  The 
global structure of the TW Hya disk has been investigated previously, using the 
SED \citep{calvet02}, optical/infrared scattered light observations 
\citep{krist00,trilling01,weinberger02,apai04,roberge05}, millimeter/radio-wave 
continuum images \citep{wilner00,wilner03,wilner05,hughes07}, and resolved 
molecular line maps \citep{qi04,qi06,qi08,hughes08,hughes11}.  However, none of 
those previous studies had the combination of angular resolution and 
sensitivity for the optically thin dust and high-quality gas tracers that are 
available from the SMA datasets presented here.  

In the following, a technique is described for extracting the structure of the 
TW Hya disk from these data using radiative transfer models.  We focus 
specifically on enabling a comparison between the radial distributions of the 
dust and CO tracers.  The approach we have adopted to make that comparison 
consists of three key steps.  First, we construct a model of the radial density 
structure of the dust that is able to reproduce our resolved observations of 
870\,$\mu$m continuum emission and the broadband SED.  Next, we make the 
assumption of a radially constant (vertically-integrated) dust-to-gas mass 
ratio and use the model structure we derive from the dust to predict the 
emission morphology of the CO $J$=3$-$2 line.  Then, we show that this 
assumption implies an inconsistency with the observations, highlighting a clear 
difference in the radial distributions of millimeter-sized dust grains and CO 
gas in the TW Hya disk.  Some of the potential implications of this 
inconsistency are discussed further in \S 5.

\subsection{Dust Structure}

The dust disk structure is determined following the technique outlined by 
\citet{andrews11}, with some modifications for generality.  We assume the dust 
is spatially distributed with a parametric two-dimensional density structure in 
cylindrical-polar coordinates \{$r$, $z$\}, 
\begin{equation}
\rho_d(r,z) = \frac{\Sigma_d}{\sqrt{2\pi}z_d} \exp{\left[-\frac{1}{2}\left(\frac{z}{z_d}\right)^2\right]},
\end{equation}
where $\Sigma_d$ and $z_d$ are surface densities and characteric heights, which 
both vary radially (see below).  As will be explained further in \S 4.3, we 
investigated two different models for the radial surface density profile.  
First, we employed the similarity solution for simple viscous accretion disk 
structures \citep{lynden-bell74} that we have used successfully to characterize 
both normal and transition disks in the past 
\citep{andrews09,andrews10a,andrews10b,andrews11,hughes10}.  In that case, 
\begin{equation}
\Sigma_d(r) = \Sigma_c \left(\frac{r}{r_c}\right)^{-\gamma} \exp{\left[-\left(\frac{r}{r_c}\right)^{2-\gamma}\right]},
\end{equation}
where $\Sigma_c$ is a normalization, $r_c$ is a characteristic scaling radius, 
and $\gamma$ is a gradient parameter.  As an alternative, we considered a less 
physically motivated (but perhaps more commonly used) model that incorporates a 
power-law density profile with a sharp cut-off \citep[see][]{andrews08},
\begin{equation}
\Sigma_d(r) = \Sigma_0 \left(\frac{r}{r_0}\right)^{-p} \,\,\,\,\,\,\, ({\rm if} \,\,\, r \le r_0; \,\, {\rm else} \,\,\, \Sigma_d = 0),
\end{equation}
where $\Sigma_0$ is a normalization, $r_0$ is the outer edge of the disk, and 
$p$ is a gradient parameter.  In either case, the surface densities at small 
radii are modified to account for the TW Hya disk cavity 
\citep{calvet02,hughes07}.  To simplify the inner disk model of 
\citet{andrews11}, we set the surface densities to a constant value 
$\Sigma_{\rm in}$ between the sublimation radius ($r_{\rm sub}$) and a ``gap" 
radius ($r_{\rm gap}$).  No dust is present between that gap radius and the 
cavity edge, $r_{\rm cav}$.  In the vertical dimension, the dust is distributed 
like a Gaussian with a variance $z_d^2$.  The characteristic height varies with 
radius like $z_d = z_0 (r/r_0)^{1+\psi}$.  Following \citet{andrews11}, we 
employ a cavity ``wall" to reproduce the infrared spectrum of TW Hya (no such 
feature was required at the sublimation radius).  The local value of $z_d$ is 
scaled up to $z_{\rm wall}$ at $r_{\rm cav}$, and then exponentially joined to 
the global $z_d$ distribution over a small radial width, $\Delta r_{\rm wall}$. 

This structure model has 11 parameters: three describe the base surface density 
profile, \{$\Sigma_c$, $r_c$, $\gamma$\} or \{$\Sigma_0$, $r_0$, $p$\}, five 
determine the cavity and inner disk properties, \{$\Sigma_{\rm in}$, $r_{\rm 
sub}$, $r_{\rm gap}$, $r_{\rm cav}$, $\Delta r_{\rm wall}$\}, and three others 
characterize the vertical distribution of dust, \{$z_0$, $z_{\rm wall}$, 
$\psi$\}.  To simplify the modeling, we fixed some of the parameters that are 
of less direct interest here.  The sublimation radius was set to $r_{\rm sub} = 
0.05$\,AU, the location where dust temperatures reach 1400\,K \citep[see 
also][]{eisner06}.  The gap radius was set to $r_{\rm gap} = 0.3$\,AU and the 
(constant) inner disk density to $\Sigma_{\rm in} = 5\times10^{-4}$\,g 
cm$^{-2}$.  The cavity edge was fixed at $r_{\rm cav} = 4$\,AU 
\citep[see][]{hughes07}, the wall height was set to $z_{\rm wall} = 0.25$\,AU, 
and the wall width to $\Delta r_{\rm wall} = 1$\,AU.  Since the details of this 
gap are not the focus, no attempt was made to reconcile the models with 
infrared interferometric data \citep[but see][]{eisner06,ratzka07,akeson11}.  
After extensive experimentation with modeling the SED, we also fixed the scale 
height gradient to $\psi = 0.25$.  The interplay and degeneracies between these 
free parameters were discussed in detail by \citet{andrews11}.  For our 
purposes here, it is worth emphasizing that the parameters we have fixed have 
little quantitative impact on the derived radial structures (i.e., sizes and 
density gradients).  

We used the dust composition advocated by \citet{pollack94}, consisting of a 
mixture of astronomical silicates, water ice, troilite, and organics with the 
abundances, optical properties, and sublimation temperatures discussed by 
\citet{dalessio01}.  Based on the efforts of \citet{uchida04} to faithfully 
reproduce the details of the {\it Spitzer} IRS spectrum, we let 25\%\ of the 
total silicate abundance inside the disk cavity ($r \le r_{\rm cav}$) be 
composed of crystalline forsterite \citep[using optical constants 
from][]{jager03}.  Two grain populations were employed, with a power-law size 
($s$) distribution, $n(s) \propto s^{-3.5}$, between $s_{\rm min} = 
0.005$\,$\mu$m and a given $s_{\rm max}$.  Outside the cavity wall, 95\%\ of 
the dust (by mass) has $s_{\rm max} = 1$\,mm and the remaining 5\%\ has $s_{\rm 
max} = 1$\,$\mu$m.  The dust in the wall itself and the tenuous inner disk was 
assumed to have $s_{\rm max} = 1$\,$\mu$m.  No effort was made to distinguish 
the vertical distributions of these dust populations.  Opacity spectra for each 
population were determined from Mie calculations, assuming segregated spherical 
grains.  For these dust assumptions, the 870\,$\mu$m dust opacity in the outer 
disk is $\kappa_{\rm mm} = 3.4$\,cm$^2$ g$^{-1}$.

We assumed the central star has a K7 spectral type with $T_{\rm eff} = 
4110$\,K, $R_{\ast} = 1.04$\,R$_{\odot}$, and $M_{\ast} = 0.8$\,M$_{\odot}$ 
($\log{g} = 4.3$), based on an effort to match \citet{lejeune97} spectral 
synthesis models to the broadband SED and the detailed optical/infrared 
spectral analysis work of \citet{yang05}.  The best-fit stellar spectrum 
template is overlaid on the SED in Figure \ref{data}$d$ as a light gray curve.
Recently, \citet{vacca11} have argued instead for a M2.5 spectral type in the 
near-infrared, and a correspondingly cooler stellar photosphere (3400\,K), 
larger radius (1.29\,R$_{\odot}$), and lower mass (0.4\,M$_{\odot}$).  While 
that stellar model provides a good match to the broadband infrared photometry 
for TW Hya, it underpredicts the observed optical fluxes by a factor of 
$\sim$3 \citep[in the $BVR$ bandpasses, and its known variability does not 
bridge that gap; see][]{mekkaden98}.  We prefer the parameters for the warmer 
photosphere because they produce a template spectrum that better matches the 
SED across the complete set of optical and infrared bandpasses.

For a given set of parameters, we simulated the stellar irradiation and 
emission output of a model dust structure using the two-dimensional, 
axisymmetric Monte Carlo radiative transfer code {\tt RADMC} 
\citep[see][]{dullemond04a}.  Assuming the fixed viewing geometry determined by 
\citet{hughes11}, a raytracing algorithm was then used to compute a synthetic 
model SED and set of 870\,$\mu$m continuum visibilities sampled at the same 
spatial frequencies observed with the SMA.  For each surface density model, we 
found the best simultaneous fit to the observed SED and SMA visibilities over a 
coarse grid of the gradient parameter $\gamma$ or $p$, by varying the 
parameters \{$\Sigma_c$ or $\Sigma_0$, $r_c$ or $r_0$, $z_0$\}.  Based on those 
results, we refined our search and permitted the gradients to vary freely to 
find the best-fit parameter sets for each model type (see \S 4.3 for results).

\subsection{CO Gas Structure}

Unlike for the dust, the radial density profile of the gas disk cannot be 
inferred directly from models of the optically thick $J$=3$-$2 transition of 
CO.  Therefore, we make a fundamental assumption that the gas traces the dust 
in the radial dimension.  For any given $\Sigma_d$, we define the gas surface 
density profile as $\Sigma_g = \Sigma_d/\zeta$, where $\zeta$ is a (radially) 
constant dust-to-gas mass ratio.  

\begin{figure}
\epsscale{0.5}
\plotone{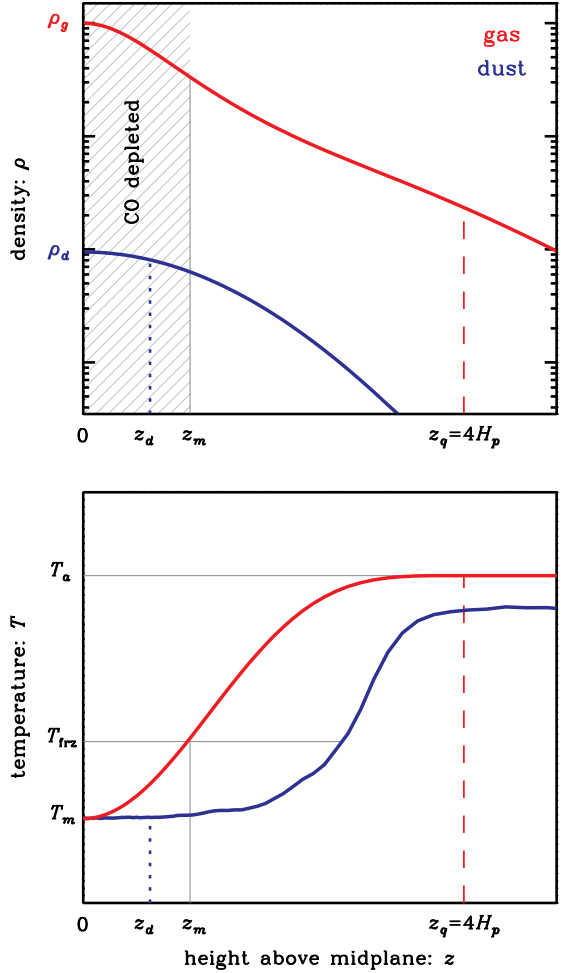}
\figcaption{Schematic demonstration of our model vertical structure, shown as
a vertical slice at a fixed radius.  ({\it top}) The density profile of the gas 
({\it red}) and dust ({\it blue}) as a function of height above the midplane.  
The latter has a parametric Gaussian distribution with a variance $z_d^2$, 
while the former is computed assuming it is in hydrostatic pressure balance for 
its specified temperature structure.  The relative normalizations of each are 
represented accurately, such that $\Sigma_d = \zeta \Sigma_g$, where $\zeta$ is 
a fixed dust-to-gas mass ratio.  The hatched region near the midplane ($z \le 
z_m$) marks where CO is depleted from the gas phase because it is frozen onto 
dust grain mantles (where $T_g \le T_{\rm frz}$).  The comparable surface CO 
depletion zone due to photodissociation ($z \ge z_s$) is well off the right of 
the plot.  ({\it bottom}) The corresponding temperature profiles, where $T_d$ 
is computed from the {\tt RADMC} radiative transfer calculations and $T_g$ is 
determined parametrically, as described in the text.  The gas and dust 
temperatures are equivalent ($T_g = T_d = T_m$) in the midplane, but $T_g$ 
rises more rapidly than the dust before it saturates to a value $T_a$ at a 
height $z_q$.  \label{model_fig}}
\end{figure}

However, we have elected to permit some freedom in the vertical distribution of 
the gas to facilitate a more faithful reproduction of the CO channel maps.  
Using a multi-transition CO dataset, \citet{qi06} noted that models of the TW 
Hya disk structure had a difficult time reproducing an appropriate vertical 
temperature gradient of the gas.  The intensity of the high-excitation 
$J$=6$-$5 line indicated that the gas in the disk atmosphere was significantly 
hotter than the dust, presumably due to substantial X-ray heating from the 
central star.  To be able to reproduce the observed CO spectral images, we have 
characterized the vertical temperature profile of the gas in parametric form, 
based on the modeling analysis of \citet{dartois03}.  We assume that
\begin{equation}
T_g(r,z) = T_a + (T_m-T_a) \cos{\left( \frac{\pi z}{2 z_q} \right)}^{2\delta},
\end{equation}
where $T_a = T_1(r/{\rm 1\,AU})^{-q}$ is a parametric radial temperature 
profile in the disk atmosphere, $T_m$ is the midplane temperature determined 
from the {\tt RADMC} simulations of the dust, $\delta$ describes the shape of 
the vertical profile, and $z_q$ defines the height of the atmosphere layer 
such that $T_g(z \ge z_q) = T_a$.  In our modeling, we fix $\delta = 2$ and 
$z_q = 4H_p$, where $H_p$ is the pressure scale height assuming the midplane 
temperature at each radius ($H_p = c_s/\Omega$, the ratio of the midplane sound 
speed to the Keplerian angular velocity).  In practice, Eq.~(4) is a reasonable 
parametric approximation of the vertical temperature profile in an irradiated 
disk \citep[e.g.,][]{dalessio99}.  We have grounded the models by forcing $T_g 
= T_d$ at the midplane ($z = 0$), but allowed the gas temperatures to increase 
faster with height than the dust to simulate any additional external heating 
sources.  For a given $T_g(r,z)$ specified by \{$T_1$, $q$\}, we then calculate 
the vertical density structure of the gas by numerically integrating the 
equation of vertical hydrostatic equilibrium,
\begin{equation}
\frac{\partial \ln{\rho_g}}{\partial z} = - \left[ \left(\frac{G M_{\ast} z}{(r^2+z^2)^{3/2}}\right)\left(\frac{\mu m_H}{k T_g}\right) + \frac{\partial \ln{T_g}}{\partial z} \right]
\end{equation}
using $\Sigma_g$ as a boundary condition, where $G$ is the gravitational 
constant, $\mu = 2.37$ is the mean molecular weight of the gas, $m_H$ is the 
mass of a hydrogen atom, and $k$ is the Boltzmann constant.  For reference, a 
vertical slice of a representative model structure is shown in Figure 
\ref{model_fig}.  

To quantify the density of CO molecules from any given gas model structure, we 
adopt the layered approach of \citet{qi08,qi11}.  Based on the detailed 
chemical calculations of \citet{aikawa06}, we define two vertical boundaries 
\{$z_m$, $z_s$\} at any given radius such that the CO mass fraction (relative 
to H$_2$) is $X_{\rm co}$ if $z_m(r) \le z \le z_s(r)$ and $10^{-4} X_{\rm co}$ 
elsewhere.  The ``midplane" boundary, $z_m$, marks the maximum height where CO 
molecules are expected to be frozen out of the gas phase and affixed to dust 
grain mantles.  In practice, $z_m$ is defined as the minimum height where $T_g 
\ge T_{\rm frz}$, where the freeze-out temperature $T_{\rm frz}$ is a parameter considered to be constant with radius.  The ``surface" boundary, $z_s$, is 
meant to represent the height where CO molecules can be photodissociated by 
X-rays or cosmic rays.  Following \citet{qi08}, we define $z_s$ such that
\begin{equation}
N_{\rm pd} = \frac{1}{\mu m_H} \int_{\infty}^{z_s} \rho_g \,\, dz,
\end{equation}   
where $N_{\rm pd}$ is a vertically-integrated H$_2$ column density that 
effectively represents the penetration depth of the photodissociating 
radiation field: $N_{\rm pd}$ is treated as a radially constant parameter.  

So, for any dust model a corresponding CO model can be characterized with a set 
of six additional parameters: two describe the abundances of dust and CO 
relative to the total gas mass, \{$\zeta$, $X_{\rm co}$\}, two others 
characterize the gas temperatures in the disk atmosphere, \{$T_1$, $q$\}, and 
the last two define the spatial distribution of CO in the gas phase, \{$T_{\rm 
frz}$, $N_{\rm pd}$\}.  For a given set of these parameters, we generate a 
two-dimensional grid of $n_{\rm co}(r,z)$ and $T_g(r,z)$ values and define a 
velocity field based on Keplerian rotation around a point mass $M_{\ast}$, 
assuming a minimal turbulent velocity line width of 10\,m s$^{-1}$ based on the 
analysis of \citet{hughes11}.  We then feed that information into the radiative 
transfer modeling code {\tt LIME} \citep{brinch10} to solve the non-LTE 
molecular excitation conditions of the model and generate a synthetic 
high-resolution model cube for the CO $J$=3$-$2 transition.  That model cube 
was then re-sampled at the velocity resolution of the data, and its Fourier 
transform was sampled at the same spatial frequencies observed by the SMA.  In 
practice, we fixed the dust-to-gas ratio based on the assumed dust composition, 
where $\zeta = 0.014$ \citep[a gas-to-dust mass ratio of 71, 
see][]{pollack94,dalessio01}.  For each dust model, we varied \{$T_1$, $q$\} 
for a coarse grid of CO abundance layer parameters \{$X_{\rm co}$, $T_{\rm 
frz}$, $N_{\rm pd}$\} to find the best available match to the SMA spectral 
visibilities.  Since we are using only a single CO transition in this 
investigation, the abundance layer parameters do not have a strong, independent 
effect on the synthetic CO visibilities.  For simplicity, we adopt the best-fit 
models where $X_{\rm co} = 2\times10^{-6}$, $T_{\rm frz} = 20$\,K, and $N_{\rm 
pd} = 10^{21}$\,cm$^{-2}$ as representative.

\begin{deluxetable}{ccccc|cc|ccc}
\tablecolumns{10}
\tablewidth{0pc}
\tablecaption{Estimated Model Parameters and Fit Results\label{params_table}}
\tablehead{
\colhead{Model} & \colhead{$\Sigma_d$(10)} & \colhead{$\gamma$, $p$} & \colhead{$r_c$, $r_0$} & \colhead{$z_d$(10)} & \colhead{$T_a$(10)} & \colhead{$q$} & \colhead{$\chi_{\rm sed}^2$} & \colhead{$\chi_{\rm cont}^2$} & \colhead{$\chi_{\rm co}^2$} \\
\colhead{} & \colhead{[g cm$^{-2}$]} & \colhead{} & \colhead{[AU]} & \colhead{[AU]} & \colhead{[K]} & \colhead{} & \colhead{} & \colhead{} & \colhead{} \\
\colhead{(1)}     & \colhead{(2)}   & \colhead{(3)}           & \colhead{(4)}      & \colhead{(5)}        & \colhead{(6)} & \colhead{(7)} & \colhead{(8)} & \colhead{(9)} & \colhead{(10)}}
\startdata
sA      & 0.29 & 1.0  & 35 & 0.62 & 104 & 0.55 & 201 & 313,914 & 319,974 \\
sB      & 0.51 & 0.5  & 43 & 0.58 & 101 & 0.54 & 204 & 304,303 & 319,989 \\
sC      & 0.28 & 0.0  & 45 & 0.57 & 98  & 0.54 & 210 & 302,132 & 320,013 \\
sD      & 0.20 & -0.5 & 42 & 0.53 & 103 & 0.53 & 216 & 304,329 & 320,024 \\
sE      & 0.14 & -1.0 & 40 & 0.50 & 102 & 0.53 & 231 & 307,785 & 320,040 \\
\hline
pA      & 0.79 & 1.5  & 77 & 0.62 & 104 & 0.54 & 205 & 309,370 & 320,056 \\
pB      & 0.46 & 1.0  & 67 & 0.60 & 99  & 0.54 & 211 & 302,747 & 320,047 \\
pC      & 0.39 & 0.75 & 60 & 0.58 & 99  & 0.54 & 215 & 301,785 & 320,044 \\
pD      & 0.31 & 0.5  & 58 & 0.58 & 100 & 0.54 & 227 & 302,570 & 320,035 \\
pE      & 0.23 & 0.0  & 51 & 0.55 & 104 & 0.54 & 249 & 306,980 & 320,068 \\
\enddata
\tablecomments{Col.~(1): Model designation, where `s' = similarity solution and 
`p' = power-law with sharp edge as defined in Eq.~2 and 3, respectively.  
Col.~(2): Dust surface density at $r = 10$\,AU.  Col.~(3): Surface density 
gradient.  Col.~(4): Characteristic scaling radius (`s' models) or outer edge
radius (`p' models).  Col.~(5): Characteristic dust height at $r = 10$\,AU.
Col.~(6): Gas temperature in the disk atmosphere at $r = 10$\,AU (see Eq.~4).
Col.~(7): Radial gradient of the atmosphere gas temperature profile.  Col.~(8):
$\chi^2$ statistic for the SED (including {\it Spitzer} IRS spectrum).
Col.~(9): $\chi^2$ statistic for the 870\,$\mu$m visibilities.  Col.~(10): 
$\chi^2$ statistic for the CO $J$=3$-$2 spectral visibilities.  The quoted 
parameter values and fit results are valid for a set of additional fixed 
parameters, including: the gradient of the dust height profile $\psi = 0.25$, 
the sublimation radius $r_{\rm sub} = 0.05$\,AU, the inner radius of the gap 
$r_{\rm gap} = 0.3$\,AU, the cavity edge $r_{\rm cav} = 4$\,AU, the cavity wall 
height $z_{\rm wall} = 0.25$\,AU, the gas temperature profile parameters 
$\delta = 2$ and $z_q = 4H_p$, the dust-to-gas ratio $\zeta = 0.014$, the 
CO/H$_2$ abundance ratio $X_{\rm co} = 2\times10^{-6}$, the CO freezeout 
temperature $T_{\rm frz} = 20$\,K, and the CO photodissociation column $N_{\rm 
pd} = 10^{21}$\,cm$^{-2}$.  Furthermore, we assume a disk inclination of 
6\degr, major axis position angle of 335\degr, and stellar mass of 
0.8\,M$_{\odot}$.}
\end{deluxetable}

\begin{figure}
\epsscale{1.0}
\plotone{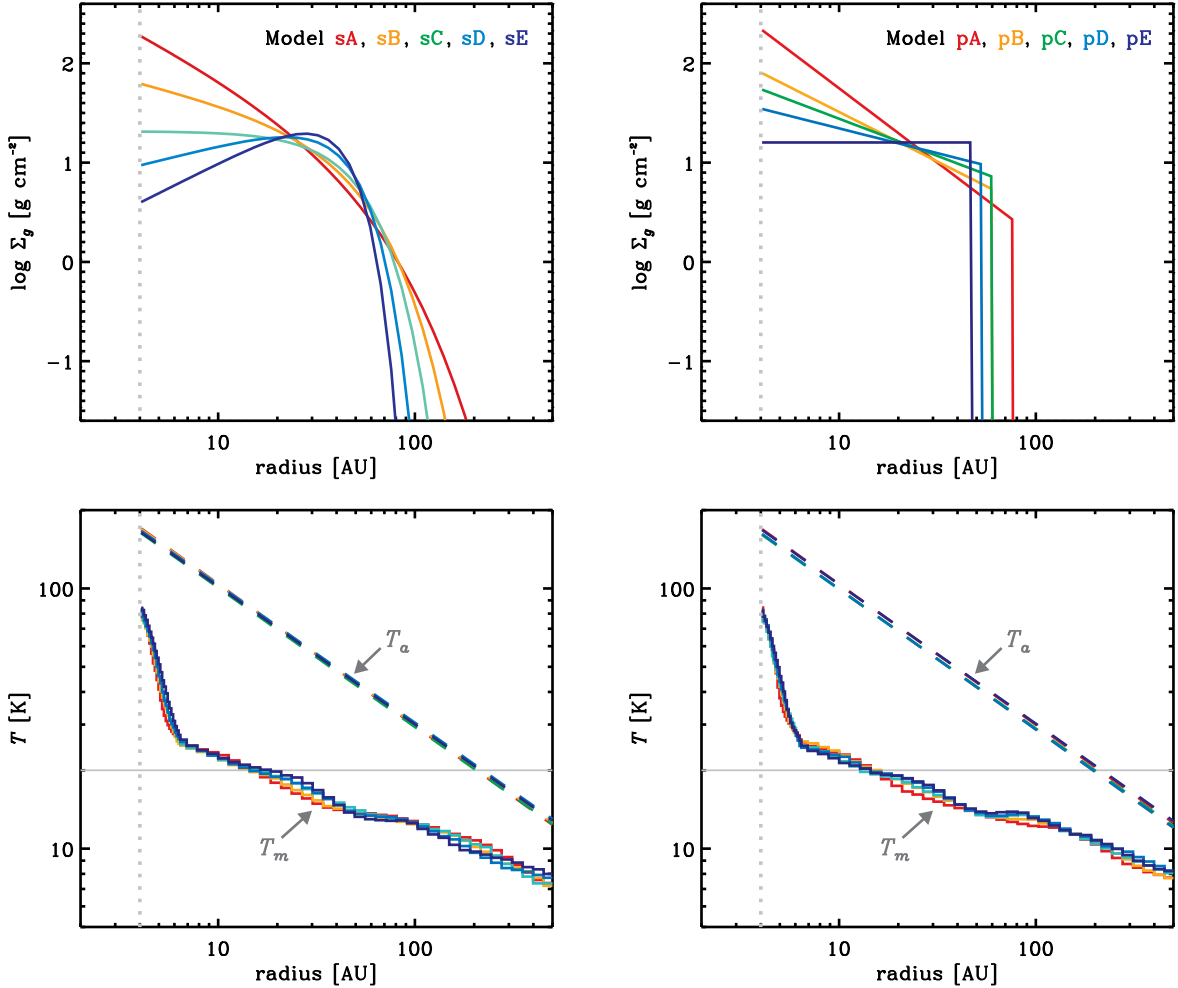}
\figcaption{({\it top}) Best-fit gas surface densities (where $\Sigma_g = 
\Sigma_d / \zeta$).  Similarity solution models are shown on the {\it left} and 
power-law + sharp edge models are shown on the {\it right}.  ({\it bottom}) 
Temperature profiles at the disk midplane ($T_m$) are shown as {\it solid} 
curves, and were determined from {\tt RADMC} radiative transfer calculations.  
The parametric atmosphere temperatures ($T_a$; at a height $z_q$, see \S 4.2)
are overlaid as dashed curves.  All models have a gap at $r = 4$\,AU, marked
with a dotted gray line.  The CO freezeout temperature, $T_{\rm frz} = 20$\,K,
is marked as a horizontal gray line.  The temperature profiles shown here are
similar for all models: the midplane values do not change much because the
total irradiated dust mass is roughly the same in each case, and the atmosphere
temperatures are determined from the same CO emission data.  However, each
model does have slight variation in the vertical temperature profile, which is
not displayed here for the sake of clarity.  \label{structures}}
\end{figure}

\begin{figure}
\epsscale{0.9}
\plotone{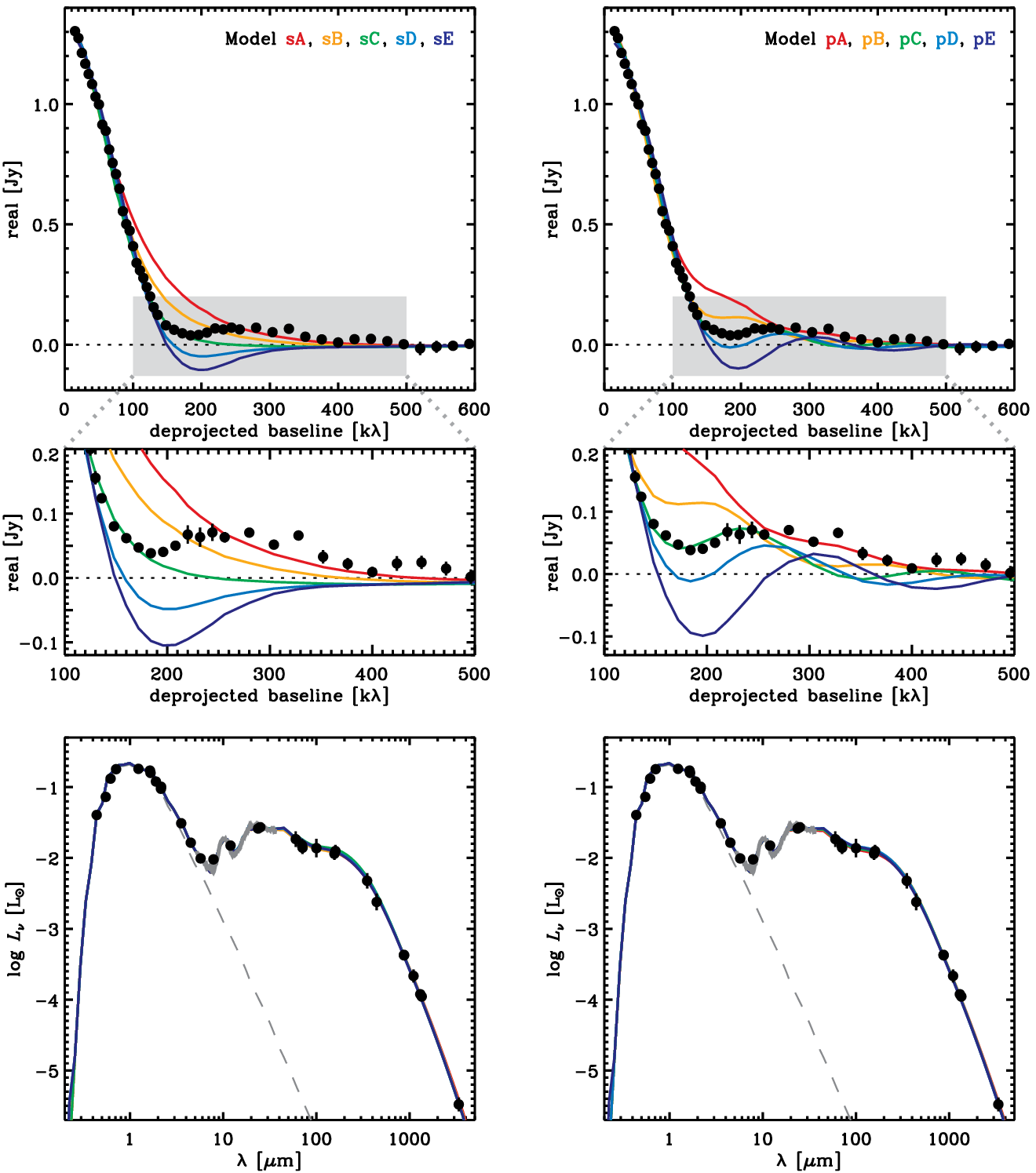}
\figcaption{A comparison of the model structures in Table \ref{params_table}
with observations of the dust continuum emission from the TW Hya disk,
including the 870\,$\mu$m visibility profile ({\it top}, with a zoomed-in view 
of the emission on longer baselines in the {\it middle}) and SED ({\it
bottom}).  The similarity solution surface density models (Eq.~2) are shown on
the left, and the power-law models with sharp outer edges (Eq.~3) are shown on
the right.  The model SEDs are essentially indistinguishable, but there are
clear differences in the 870\,$\mu$m radial emission profiles.  The overall
best match to the continuum emission is Model pC, which has a density gradient
$p = 0.75$ and a sharp outer edge at $r_0 = 60$\,AU (for the similarity
solution models, the best match is Model sC).   \label{continuum}}
\end{figure}

\subsection{Modeling Results}

The best-fit parameters for a range of representative surface density gradients 
of each model type are compiled in Table \ref{params_table}.  For clear 
notation, each model is labeled with a letter designation corresponding to its 
$\gamma$ or $p$ value, preceded by a lower-case `s' for the similarity solution 
models (based on Eq.~2) or `p' for the power-law models (based on Eq.~3).  The 
surface densities, characteristic dust heights, and gas atmosphere temperatures 
are listed for a radius of 10\,AU, where the two surface density model types 
have comparable behavior.  The four parameters describing the dust densities, 
\{$\Sigma_d(10)$, $\gamma$ or $p$, $r_c$ or $r_0$, $z_d(10)$\}, were determined 
from joint fits to the SED and 870\,$\mu$m continuum visibilities.  With those 
parameters fixed, the gas temperature parameters \{$T_a(10)$, $q$\} were 
estimated from the CO $J$=3$-$2 visibilities only.  The individual $\chi^2$ 
values for the SED, continuum visibilities, and CO visibilities are listed in 
Cols.~(8)-(10).  There are 77 independent datapoints in the SED (including 45 
equally-spaced points across the {\it Spitzer} IRS spectrum), 166,796 continuum 
visibilities, and 83,740 CO visibilities in each of 15 spectral channels (a 
total of 1,256,100).  The $\chi^2$ values were calculated with weights that 
incorporate the quadrature sum of the formal uncertainties and absolute 
calibration uncertainties on each datapoint (e.g., a 10\%\ systematic 
uncertainty on the amplitudes).  Figure \ref{structures} shows the radial 
structures for each of the disk models in Table \ref{params_table}, including 
their surface densities ({\it top}) and temperature profiles ({\it bottom}).  

Figure \ref{continuum} directly compares the 870\,$\mu$m visibilities and SEDs 
for these model structures with the observations.  All of the model structures 
provide excellent fits to the broadband SED and {\it Spitzer} IRS spectrum.  
Individual SED model behaviors are indistinguishable, aside from small 
deviations near the far-infrared turnover where the measurement uncertainties 
are largest.  However, the different structures and model types exhibit 
distinctive signatures in their resolved 870\,$\mu$m emission profiles.  The 
similarity solution models with positive density gradients ($\gamma > 0$; 
Models sA and sB) substantially over-predict the emission on 
$\sim$100-200\,k$\lambda$ scales, while those with negative gradients ($\gamma 
< 0$; Models sD and sE) have visibility nulls at $\sim$150\,k$\lambda$ that are 
clearly not commensurate with the data.  With an intermediate $\gamma = 0$, 
Model sC provides the best match to the continuum data for this model type.  
However, it and all other similarity solution models fail to reproduce the 
visibility oscillations beyond 200\,k$\lambda$.   This ``ringing" is a classic 
sign of a sharp edge in the radial emission profile, which is naturally 
produced with the power-law models described by Eq.~(3).  For those structures, 
steep density gradients ($p = 1.0$-1.5, Models pA and pB) produce too much 
emission on 100-200\,k$\lambda$ baselines and shallow gradients ($p = 0.0$-0.5, 
Models pD and pE) generate visibility nulls that are not observed.  However, an 
intermediate case with $p = 0.75$ (Model pC) is an excellent match to the 
data, with only a small (albeit significant) departure on 280-380\,k$\lambda$ 
scales.  That mismatch is likely related to the shape of the outer edge, 
although we have not pursued that speculation further.  Of all the dust 
structures explored here, Model pC is clearly favored.  

\begin{figure}
\epsscale{1.0}
\plotone{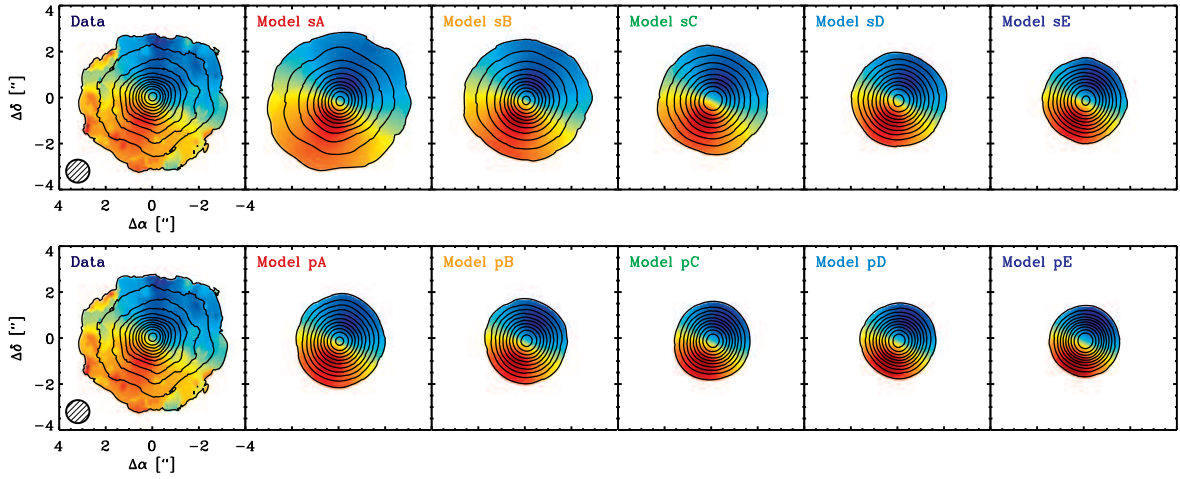}
\figcaption{Moment maps of the CO $J$=3$-$2 emission from the TW Hya disk and
the various disk structure models compiled in Table \ref{params_table}.  The
leftmost panels show the SMA observations.  The {\it top} panels make a direct
comparison with the similarity solution models, and the {\it bottom} panels do
the same for the power-law models with sharp edges.  In all plots, contours
mark the velocity-integrated CO intensities (0$^{\rm th}$ moment) at 0.4\,Jy km
s$^{-1}$ ($\sim$3\,$\sigma$) intervals and the color scale corresponds to the
intensity-weighted line velocities (1$^{\rm st}$ moment).  Only Model sA
provides a good match to the observed CO emission; all others predict gas
distributions that are too small relative to observations.
\label{moments}}
\end{figure}

\begin{figure}
\epsscale{1.0}
\plotone{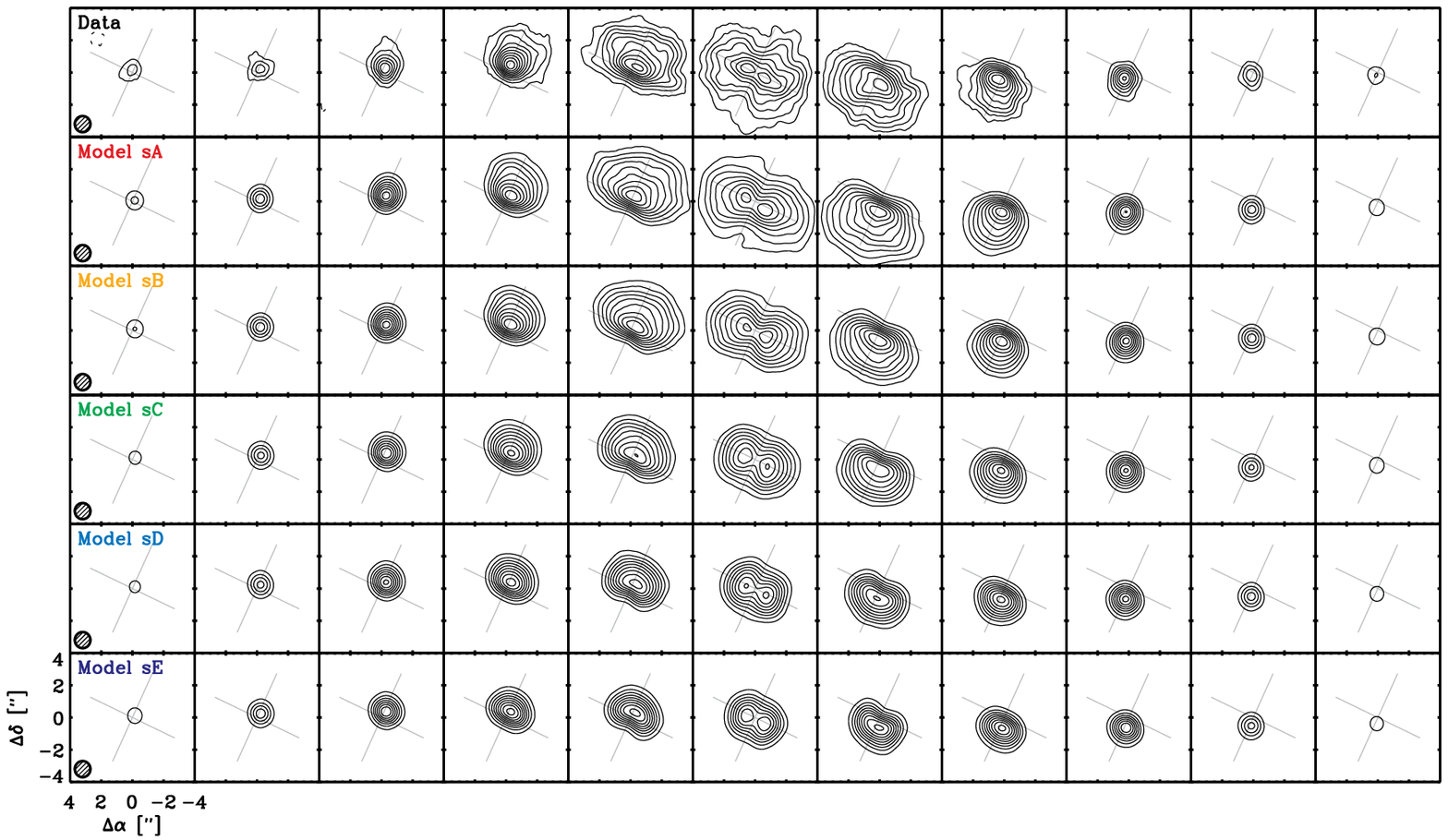}
\plotone{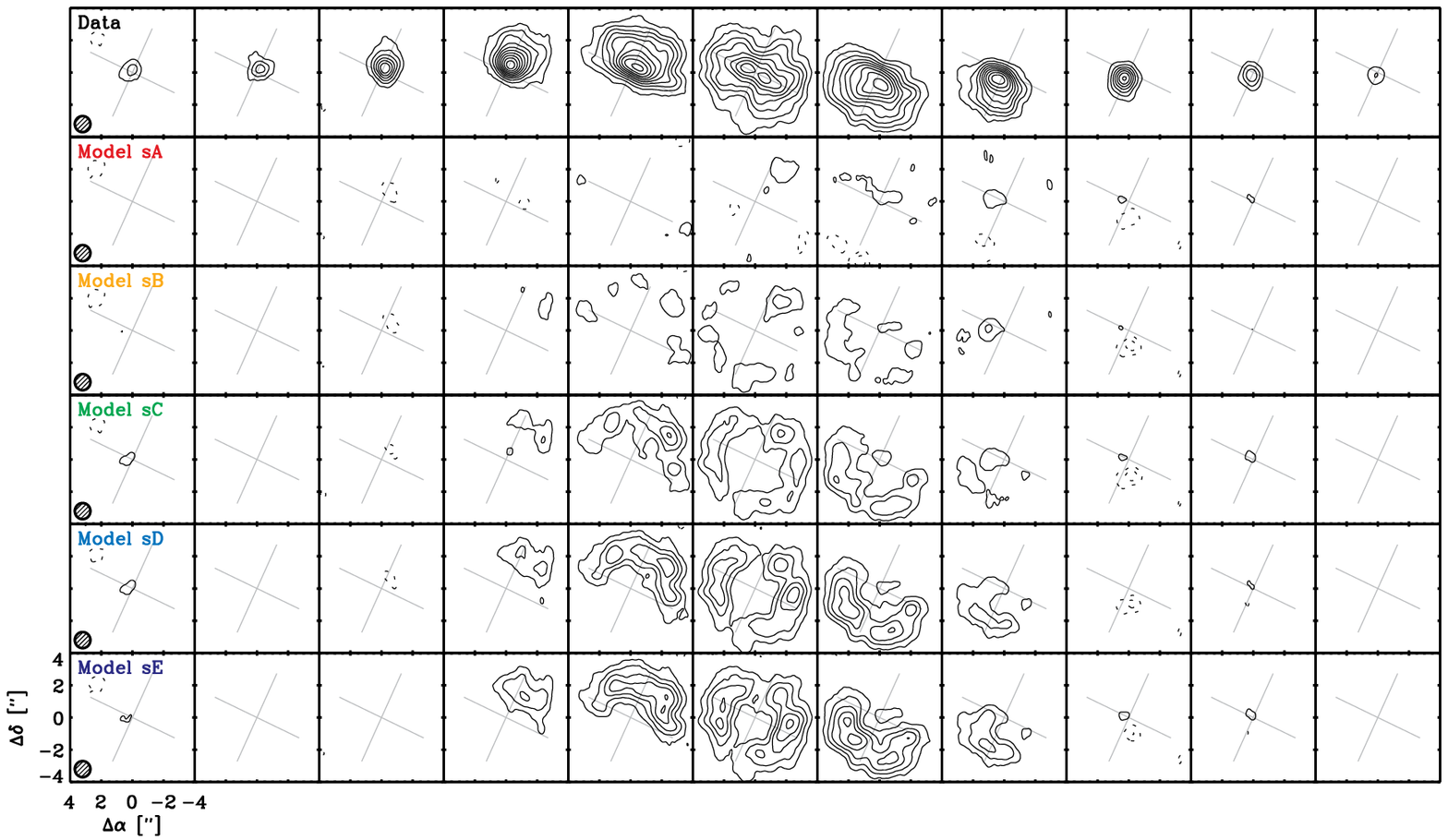}
\figcaption{A direct comparison of the observed CO $J$=3$-$2 channel maps with
synthetic data from the similarity solution models in Table
\ref{params_table}.  The {\it top} block of panels show the model spectral
visibilities synthesized in the same way as the data, and the {\it bottom}
block displays the imaged residual visibilities.  All maps have the same
contour intervals as in Figure \ref{data}$d$.  The viewing geometry of the TW
Hya disk is marked with a gray cross in each channel map.\label{chmaps_s}}
\end{figure}

\begin{figure}
\epsscale{1.0}
\plotone{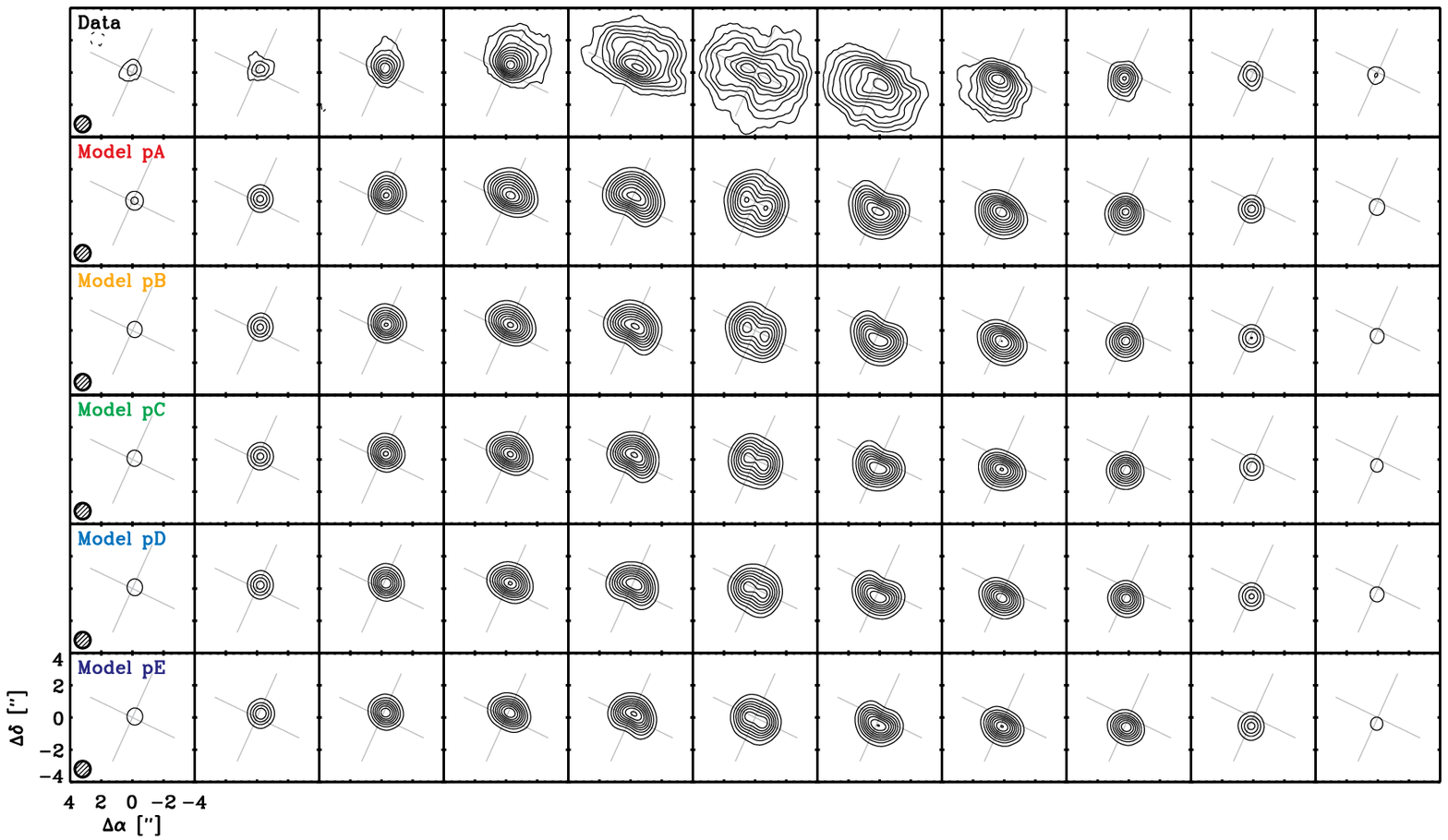}
\plotone{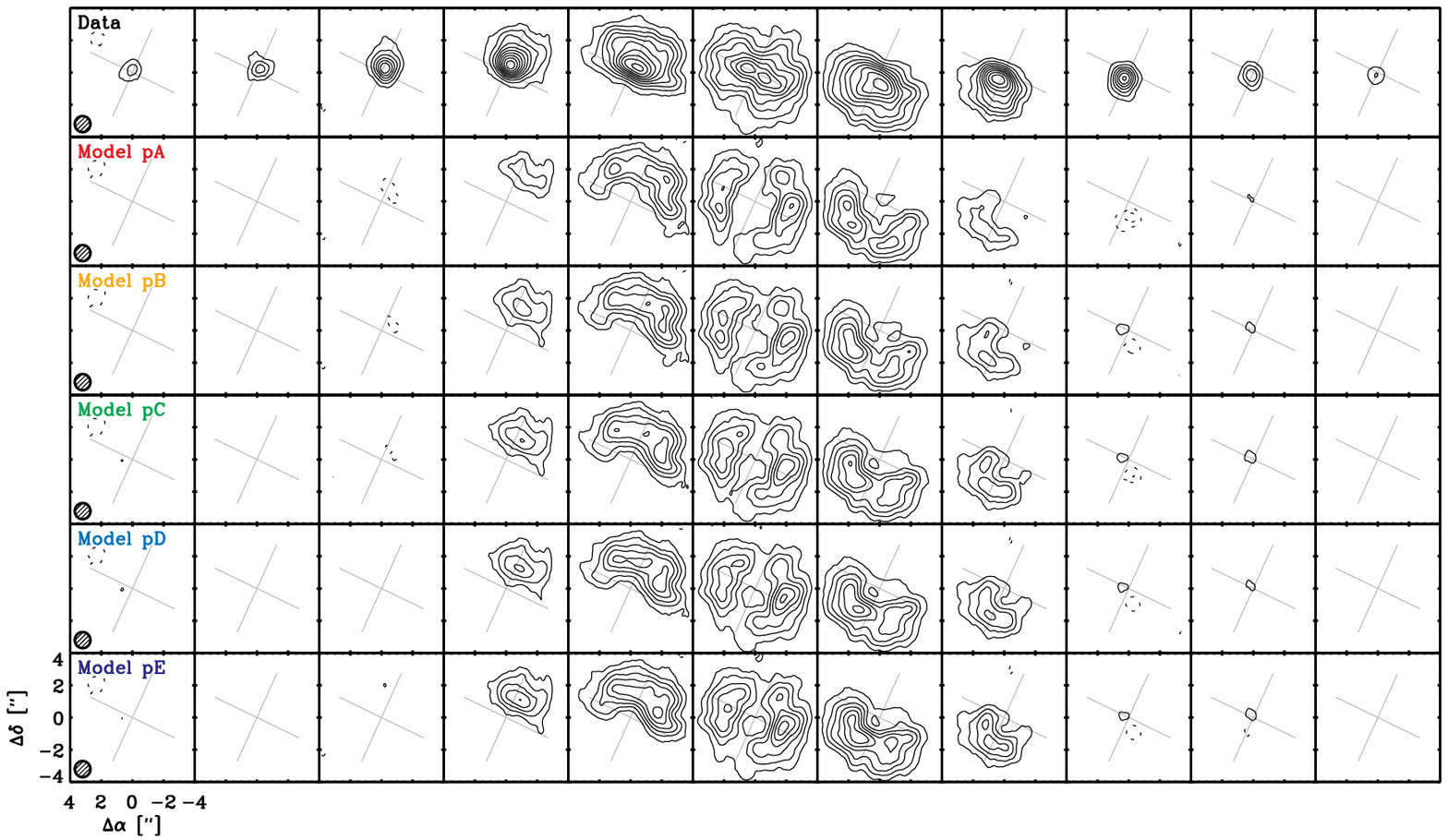}
\figcaption{Same as Figure \ref{chmaps_s}, but for the power-law models with
sharp outer edges.  \label{chmaps_p}}
\end{figure}

A consistent feature of all the dust-based model structures is their 
compactness.  The similarity solution models require characteristic radii of 
$r_c = 35$-45\,AU and the power-law models call for sharp outer edges at $r_0 = 
51$-77\,AU.  That compact dust distribution was noted in \S 3, where we 
pointed out that {\it all} of the 870\,$\mu$m dust continuum emission is 
concentrated inside a radius of roughly 1\arcsec\ ($\sim$60\,AU; see Figure 
\ref{data}$a$).  Given the much larger radial extent of the CO $J$=3$-$2 
emission -- out to radii of at least 4\arcsec\ (215\,AU; see Figure 
\ref{data}$d$) -- the best-fit models face major problems reconciling 
the CO and dust observations.  The moment maps in Figure \ref{moments} confirm 
this tension between the dust-based structure models and CO data, highlighting 
a clear CO-dust size discrepancy in nearly all cases.  More direct comparisons 
of the CO channel maps with each model structure can be made in Figures 
\ref{chmaps_s} and \ref{chmaps_p}, which show both the models ({\it top} 
panels) and the imaged residual visibilities ({\it bottom} panels) 
synthesized in the same way as the data.  Thanks to the freedom afforded by the 
parametric treatment of the gas temperatures, all models successfully reproduce 
the central CO emission core that is generated in the line wings.  However, 
only Model sA has a sufficient amount of mass at large disk radii to account 
for the observations near the systemic velocity.  Unfortunately, Model sA 
provides a poor match to the resolved 870\,$\mu$m continuum emission profile.  

Before investigating potential physical explanations for the apparent CO-dust 
size discrepancy, it is important to evaluate the possibility that this is an 
artifact of some assumption in the modeling (aside from the underlying 
parametric models themselves; see \S 5).  Admittedly, we adopted a relaxed 
approach for treating the relative vertical distributions of gas and dust in 
these models.  Only two basic requirements were enforced for consistency: the 
gas and dust are in thermal equilibrium at the disk midplane, and the gas is 
distributed vertically according to hydrostatic pressure equilibrium.  There 
are more sophisticated treatments of the gas in disk atmospheres that 
accurately account for the effects that X-ray heating, chemical 
differentiation, and gas-dust temperature departures have on the vertical 
structure of the disk \citep[e.g.,][]{vanzadelhoff01,jonkheid04,kamp04,kamp10,woitke09,woitke10,aresu11}.  A nearby disk like TW Hya would benefit from a more 
detailed analysis with such models, particularly using multiple resolved 
molecular lines \citep[see][for a start]{thi10}.  With a more physically 
motivated treatment of the vertical structure, we might infer a quantitatively 
different set of atmosphere properties that would affect the normalization and 
shape of the CO emission profile (currently controlled by \{$T_1$, $q$\} and 
other fixed parameters; see \S 4.2).  However, the added complexity of those 
models would not change the fact that a compact density distribution is 
required to explain the dust emission.  And since we assumed that the CO gas 
traces the dust in the radial direction, the size discrepancy would remain: 
these dust-based density profiles do not have enough material to be able to 
excite CO emission lines at large disk radii (see Figure \ref{structures}).  

To summarize, the fundamental conclusion from the radiative transfer modeling 
analysis of the SMA data is that the spatial distributions of the CO and 
millimeter-sized dust in the TW Hya disk are different: the dust is 
systematically inferred to be more compact than the CO gas.

\section{Discussion}

We used the SMA to observe the 870\,$\mu$m continuum and CO $J$=3$-$2 line 
emission from the disk around the nearby young star TW Hya.  These data 
represent the most sensitive high spatial resolution (down to scales of 16\,AU) 
probes of the millimeter-wave dust and molecular gas content in any 
circumstellar disk to date.  Along with the SED, these observations were used 
in concert with continuum and line radiative transfer calculations in an effort 
to extract the disk density structure.  We were unable to identify a consistent 
model structure that simultaneously accounts for the observed radial 
distributions of CO and dust.  Assuming a radially constant grain size 
distribution and (vertically integrated) gas-to-dust mass ratio, the 
millimeter-sized dust structure is significantly more compact than the CO.  The 
resolved continuum emission profile demonstrates that the radial distribution 
of the millimeter-sized solids in the TW Hya disk has a relatively sharp outer 
edge near 60\,AU, which is considerably smaller than the observed extent of the 
CO emission (out to at least 215\,AU).

The TW Hya disk structure has been studied extensively with millimeter-wave 
observations at lower angular resolution.  In a series of investigations 
focused almost exclusively on spectral line emission, \citet{qi04,qi06,qi08} 
and \citet{hughes11} relied on a slightly modified version of the physical 
structure model that was developed by \citet{calvet02} to match the TW Hya 
SED.  While that model has been used quite successfully to explain the 
molecular gas structure, there has always been some tension between 
observations and its predicted millimeter/radio continuum emission profile 
\citep[see][]{qi04,wilner03,wilner05}.  Given the modest quality of the 
previous continuum data and the fact that this structure model was not designed 
(or fitted) with access to resolved observations of any kind, the disagreement 
was understandably dismissed.  As might be expected from its assumption of a 
``steady" accretion disk structure (i.e., with a large, positive surface 
density gradient), the \citet{calvet02} model exhibits the same type of 
behavior as Models sA/B or pA/B: it agrees with the data on large scales 
($<$100\,k$\lambda$), but significantly over-predicts the amount of emission on 
smaller scales.  The same is true for the parametric similarity solution models 
(where $\gamma \approx 1$) explored by \citet{hughes08,hughes11}, and would 
certainly apply to the analogous models developed by \citet{thi10} and 
\citet{gorti11}.  All of this previous modeling work admirably reproduces the 
extended molecular line emission that is observed, and in many cases the SED 
and millimeter-wave continuum emission on large angular scales as well.  It is 
only with the sensitive, high angular resolution data presented here that we 
recognize a problem: the radial distribution of the millimeter-sized dust 
grains is much more compact than for the CO.

Unlike the thermal radiation at millimeter wavelengths, the optical and 
near-infrared light that scatters off small ($\le$1\,$\mu$m) grains in the TW 
Hya disk surface is detected out to large distances from the central star -- at 
least $\sim$4\arcsec, comparable to what is inferred from CO spectral images 
\citep{krist00,trilling01,weinberger02,apai04,roberge05}.  So {\it some} dust 
traces the molecular gas, even if it is only a limited mass of small grains up 
in the disk atmosphere.  However, these exquisitely detailed scattered light 
images exhibit subtle structural complexities.  \citet{krist00} identified four 
distinct radial zones in the optical scattered light disk, with a prominent 
steepening of the brightness distribution just outside a radius of 50\,AU 
($\sim$1\arcsec; their Zone 1/2 boundary).  Similar infrared behavior is noted 
in the studies by \citet{weinberger02} and \citet{apai04}, which both suggested 
a break in the emission profile in the 50-80\,AU ($\sim$1.0-1.5\arcsec) range.  
Those results were confirmed in a comprehensive analysis of new data by 
\citet{roberge05}, who also called attention to a color change at a similar 
radius as well as an azimuthal asymmetry out to a slightly larger distance from 
the star ($\sim$135\,AU).  The physical origin of these scattered light 
features has been a mystery, although speculation centered around variations in 
the dust height (shadowing) and gradients in the dust scattering properties 
(either mineralogical or size-related).  But in light of our discovery of an 
abrupt drop in the millimeter-wave continuum emission at the same location as 
these features, it is only natural to suspect that a more fundamental change 
occurs in the physical structure of the TW Hya dust disk near 60\,AU.

Perhaps the most straightforward explanation of the apparent CO-dust size 
discrepancy inferred from the SMA data is that we have used an incomplete 
description of the disk structure.  As an example, consider a modification of 
either model type that incorporates an abrupt decrease in the surface densities 
(or millimeter-wave dust opacities) -- not the dust-to-gas ratio -- outside $r 
\approx 60$\,AU.  If that drop in $\Sigma$ (or $\kappa_{\rm mm}$) was not too 
large (perhaps a factor of $\sim$100), there would still be enough disk 
material to produce bright emission from the optically thick CO lines and 
scattered light while also accounting for the sharp edge feature noted in the 
optically thin 870\,$\mu$m emission profile.  That ``substructure" in the outer 
dust disk might actually enhance the local gas-phase CO abundance, as 
ultraviolet radiation can penetrate deeper into the disk interior and 
photodesorb CO from the (small) reservoir of cold dust grains that remains at 
large radii \citep[e.g.,][]{hersant09}.  

Nevertheless, a physical origin for such a dramatic drop in the dust densities 
and/or the millimeter-wave dust opacities is not obvious.  One possibility is 
that the disk has been perturbed by a long-period (as yet unseen) companion.  
If a faint object is embedded in the disk near the apparent edge of the 
870\,$\mu$m emission distribution, it might open a gap that splits the disk 
into two distinct reservoirs and generate the warp asymmetry suggested by 
\citet{roberge05}.  But, a narrow gap alone would not account for the SMA 
continuum observations.  The millimeter-wave luminosity exterior to the gap 
would still need to be decreased, perhaps because the particles at those larger 
radii were preferentially unable to grow to millimeter sizes.  
\citet{weinberger02} quote deep limits on $H$-band point sources in the TW Hya 
disk that suggest there are no companions more massive than $\sim$6\,M$_{\rm 
Jup}$ near 60\,AU, according to the \citet{baraffe03} models \citep[the 
corresponding mass limit is higher for the models of][]{marley07}.  Certainly 
this kind of truncation or other forms of substructure could be invoked to 
explain the sharp radial edge in the SMA dust observations.  But rather than 
engage in further speculation on the details, it should suffice to point out 
that the potential for substructure or other anomalies in the TW Hya disk can 
be tested with a substantial increase in resolution and sensitivity.  Moreover, 
spectral imaging of optically thinner gas tracers (e.g., the CO isotopologues) 
would make for an ideal test of the origins of the apparent CO-dust size 
discrepancy.  Fortunately, such observations will shortly be available as the 
Atacama Large Millimeter Array (ALMA) begins routine science operations.  

There is a compelling alternative explanation that has a more concrete physical 
motivation.  In any protoplanetary disk, the thermal pressure of the gas is 
thought to cause it to orbit the star at slightly sub-Keplerian rates, 
generating a small velocity difference relative to the particles embedded in it 
\citep{weidenschilling77}.  Depending on their size, particles can experience a 
head-wind from this gas drag that decays their orbits and sends them spiraling 
in toward the central star.  This {\it radial drift} of particles modifies the 
radial dust-to-gas mass ratio profile and introduces a pronounced spatial 
gradient in the particle size distribution -- in essence, it causes $\zeta$ and 
$\kappa_{\rm mm}$ to decrease with radius 
\citep{brauer07,brauer08,birnstiel09}.  In the outer disk, the drift rates are 
expected to be largest for millimeter-sized particles 
\citep[e.g.,][]{takeuchi02}.  Since thermal emission peaks at a wavelength 
comparable to the particle size, the drift process should then naturally 
produce a millimeter-wave emission profile that is considerably more compact 
than would be inferred from tracers of the gas \citep{takeuchi05}.  Moreover, 
the dust particles that reflect light in the disk atmosphere are small enough 
to be dynamically coupled to the gas -- therefore, scattered light images 
should be extended like the probes of the gas phase.  

From a qualitative perspective, our analysis of the CO and dust structures in 
the TW Hya disk are certainly consistent with a scenario where growth and 
radial drift have had an observable impact.  However, it is unclear if 
realistic models of the growth and migration of solids in a disk like this can 
quantitatively account for the details.  One particular challenge worth 
highlighting is related to timescales.  The \citet{takeuchi05} models suggest 
that millimeter-wave dust emission should be strongly attenuated on a timescale 
of $\sim$1\,Myr without constant replenishment (presumably from growth and/or 
fragmentation).  Given the advanced age of TW Hya ($\sim$8-20\,Myr), this model 
would require either that replenishment shuts off after several Myr or that 
drift is inefficient at early times before becoming more important later in the 
disk evolution process.  If this is indeed the mechanism responsible for our 
results, additional observations of disks at a range of ages could be used to 
help calibrate models of the long-term evolution of drift rates.  One potential 
way to test this hypothesis relies on the particle size dependence of the 
radial drift rates.  The theory implies that larger particles will end up with 
more centrally concentrated density distributions.  At long and optically thin 
wavelengths, we expect that the size of the continuum emission region should be 
anti-correlated with wavelength -- longer wavelengths imply more compact 
emission.  In principle, this could be tested in the near future by combining 
high resolution ALMA and Expanded Very Large Array (EVLA) observations of the 
TW Hya dust disk that span the millimeter/radio spectrum.  

In many ways, TW Hya and its disk are unique and may not be representative of 
the bulk population of pre-main sequence stars and their circumstellar 
material.  Nevertheless, it is worth considering the broader implications of 
our findings for this specific example -- they {\it may} prove to be more 
generally applicable.  It is possible that the CO-dust size discrepancy found 
here is present in most other millimeter-wave disk observations, but it would 
likely be difficult to identify with current sensitivity and resolution 
limitations.  If that feature is common and its underlying cause is a drop in 
the dust-to-gas ratio in the outer disk, there would be serious consequences 
for disk mass estimates based on dust continuum measurements.  There could be 
large and hidden mass reservoirs of molecular gas in the outer reaches of 
protoplanetary disks, implying that disk masses might be substantially 
under-estimated.  If true, gas densities at large disk radii may be higher than 
typically assumed, with profound implications for facilitating giant planet 
formation by gravitational instability 
\citep[e.g.,][]{boley09,kratter10,boss11}.  However, if radial drift is the key 
process responsible for the apparent CO-dust size discrepancy, it is also 
possible that any ``primordial" dust-to-gas ratio {\it integrated over the 
entire disk} is preserved.  This would be the case if the millimeter-sized dust 
originally present at large radii had its inward migration halted before it was 
accreted onto the central star.  In that scenario, the total disk mass 
estimates from millimeter-wave luminosities would still be relatively accurate 
(assuming a proper model for the dust opacities that considers the simultaneous 
particle size evolution is available), although the densities in the outer disk 
would remain uncertain without measurements of optically thin gas emission 
lines.

Independent of the apparent CO-dust size discrepancy, our finding that the 
millimeter-wave continuum emission from the TW Hya disk is so sharply 
truncated comes as a surprise.  Modern interferometric datasets generally do 
not have sufficient sensitivity to differentiate between dust density models 
with sharp edges or smooth tapers \citep[e.g., see][]{isella10,guilloteau11}.  
Given that ambiguity, it is possible that the edge feature we have identified 
in the 870\,$\mu$m emission profile of the TW Hya disk could be relatively 
common.  Moreover, it is tempting to associate this $r \approx 60$\,AU edge in 
the distribution of larger solid particles in the TW Hya disk with the 
similarly abrupt truncation of the classical Kuiper Belt at $r \approx 
40$-50\,AU in our solar system \citep{trujillo01,gladman01}.  If this behavior 
ends up being a generic feature in protoplanetary disks, it may signify an 
important diagnostic of the radial migration of disk solids and provide new 
insights into the structural origins and evolution of the outer solar system 
\citep[e.g., see][]{kenyon99,levison03}. 

Further speculation on the generality of the features identified in the TW Hya 
disk is unnecessary.  With the recent start of ALMA science operations, the 
quality of the data presented here will be matched (and exceeded) routinely for 
large samples, and the basic trends of disk properties like those probed here 
will be clarified.  If the TW Hya disk is not anomalous, it is clear that the 
general methods used to interpret observations of dust in disks will need to be 
modified to focus less on their bulk density structures and more on the 
dynamical evolution of their solid contents.

\section{Summary}

We have presented sensitive, high resolution ($0\farcs3 = 16$\,AU) SMA 
observations of the 870\,$\mu$m continuum and CO $J$=3$-$2 line emission from 
the disk around the nearby young star TW Hya.  Based on two different 
parametric formulations for the disk densities, we used radiative transfer 
calculations to compare the predicted radial structures of the dust and CO gas 
in the TW Hya disk with these SMA observations and ancillary measurements of 
the spectral energy distribution.  The key conclusions from this modeling 
analysis of these high-quality data include:

\begin{enumerate}
\item Under the assumption that the dust-to-gas surface density ratio is 
constant with radius, we were not able to find any model structure that can 
simultaneously reproduce the resolved brightness profiles of the 870\,$\mu$m 
continuum and CO line emission.  We have identified a clear CO-dust size 
discrepancy that is present regardless of whether the assumed surface density 
profile has a sharp outer edge or a smooth (exponential) taper at large radii.

\item The radial distribution of millimeter-sized dust grains in the TW Hya 
disk is substantially more compact than its CO gas reservoir.  The 870\,$\mu$m 
dust emission has a {\it sharp} outer edge near 60\,AU, while the CO emission 
(and optical/infrared scattered light from small grains that are dynamically 
coupled to the gas) extends to a radius of at least 215\,AU.  

\item The observationally inferred CO-dust size discrepancy could potentially 
be explained with a more complex dust density profile that exhibits a sudden 
decrease by a large factor near a radius of 60\,AU.  That ``break" in the 
density profile might be consistent with a tidal perturbation by a long-period 
(unseen) companion, although the dust at larger radii would then also have to 
be preferentially depleted of millimeter-sized grains.  

\item Alternatively, the observations might have uncovered some preliminary 
evidence for a key evolutionary mechanism related to the planet formation 
process: the growth and inward migration of disk solids.  The radial drift of 
millimeter-sized particles is expected to naturally concentrate long-wavelength 
thermal emission near the star relative to tracers of the molecular gas 
reservoir.  However, a more detailed exploration of disk evolution models is 
needed to verify if the observations of the TW Hya disk are quantitatively 
consistent with this scenario.

\end{enumerate}

\acknowledgments We thank Mark Gurwell for his assistance with the multi-epoch 
calibration of the SMA data, Elise Furlan and the {\it Spitzer} IRS Disks team 
for providing a reduced TW Hya spectrum, and an anonymous referee for 
thoughtful suggestions that contributed significant value to the article.  We 
are especially grateful to Christian Brinch for his generous help with the {\tt 
LIME} code, Kees Dullemond for technical support with the {\tt RADMC} package, 
and Paola D'Alessio for her valuable advice on dust populations and optical 
constants.  The Submillimeter Array (SMA) is a joint project between the 
Smithsonian Astrophysical Observatory and the Academia Sinica Institute of 
Astronomy and Astrophysics and is funded by the Smithsonian Institution and the 
Academia Sinica.

\end{document}